# Fast, label-free tracking of single viruses and weakly scattering nanoparticles in a nano-fluidic optical fiber

**Authors:** Sanli Faez[1§], Yoav Lahini[2,3§], Stefan Weidlich[4,5], Rees F. Garmann[6], Katrin Wondraczek[5], Matthias Zeisberger[5], Markus A. Schmidt[5,7], Michel Orrit[1], Vinothan N. Manoharan[6,3*]

**Affiliations:**

[1] Huygens-Kamerlingh Onnes Laboratory, Leiden University, P.O. Box 9504, 2300 RA Leiden, The Netherlands

[2] Department of Physics, Massachusetts Institute of Technology, Cambridge, MA, USA.

[3] Department of Physics, Harvard University, Cambridge, MA, USA

[4] Heraeus Quarzglas GmbH & Co. KG, Hanau, Germany

[5] Leibniz Institute of Photonic Technology, Albert-Einstein-Str. 9, 07754 Jena, Germany

[6] Harvard John A. Paulson School of Engineering and Applied Sciences, Harvard University, Cambridge, MA, USA

[7] Otto Schott Institute of Material Research, Fraunhoferstr. 6, 07743 Jena, Germany

[§] These authors contributed equally

[*] Correspondence to: vnm@seas.harvard.edu

## Abstract

High-speed tracking of single particles is a gateway to understanding physical, chemical, and biological processes at the nanoscale. It is also a major experimental challenge, particularly for small, nanometer-scale particles. Although methods such as confocal or fluorescence microscopy offer both high spatial resolution and high signal-to-background ratios, the fluorescence emission lifetime limits the measurement speed, while photobleaching and thermal diffusion limit the duration of measurements. Here we present a tracking method based on elastic light scattering that enables long-duration measurements of nanoparticle dynamics at rates of thousands of frames per second. We contain the particles within a single-mode silica fiber containing a sub-wavelength, nano-fluidic channel and illuminate them using the fiber's strongly confined optical mode. The diffusing particles in this cylindrical geometry are continuously illuminated inside the collection focal plane. We show that the method can track unlabeled dielectric particles as small as 20 nm as well as individual cowpea chlorotic mottle virus (CCMV) virions – 4.6 megadaltons in size – at rates of over 3 kHz for durations of tens of seconds. Our setup is easily incorporated into common optical microscopes and extends their detection range to nanometer-scale particles and macromolecules. The ease-of-use and performance of this technique support its potential for widespread applications in medical diagnostics and micro total analysis systems.

**Introduction**

Measuring the dynamics of nanoparticles inside biomolecular solutions and mixtures reveals critical information about molecular-scale interactions [1], self-assembly pathways [2], the composition of the local environment [3], and the activity of components such as molecular motors [4, 5]. To make such measurements, one must not only detect the nanoparticles, but also track their motion in real space. Whereas detection involves determining the existence of a particle – and possibly measuring its size or position – tracking involves repeatedly measuring its position on timescales shorter than the characteristic diffusion or advection time. The spatial precision of these position measurements should be comparable to molecular scales, particularly when the particles are used as probes of the local environment. Despite many recent advances in detection [6, 7, 8] and optical nanoscopy [9, 10], a method with sufficiently high spatial and temporal precision to track weakly scattering and rapidly diffusing nanoparticles such as liposomes or viruses is still missing.

Tracking such nanoparticles is difficult because it requires both high measurement rates and long observation times to obtain good statistics on the motion. In fluorescence microscopy and associated techniques, the most commonly used methods in biophysical tracking experiments, the measurement rate is limited by the inherently low fluorescent emission rate, and the longest observation time is limited by photo-bleaching. Even in the absence of bleaching, the observation time is limited to the time the labeled particle spends in the imaging focal plane or illuminated volume. For a mobile particle in a viscous liquid such as water, this timescale decreases linearly with the particle size; it takes only a few microseconds for a 20-nm-diameter particle to diffuse over a distance equal to its own diameter. Tracking nanoparticles in free solution over long times therefore requires methods such as anti-Brownian electrostatic trapping [11]. Wang and Moerner [12] have recently pushed fluorescent tracking close to its physical limits by combining electrostatic trapping, photon-stamped positioning, and advanced machine-learning algorithms.

Detection schemes that use elastic light scattering have advantages over fluorescence methods for tracking on short timescales. Unlike fluorescence, elastic scattering is an instantaneous and energy-conserving process and thus has an almost unlimited photon budget, enabling high measurement rates. However, because the intensity of light elastically scattered from nanoparticles scales as the sixth power of the particle size, small nanoparticles can be difficult to detect over the background scattering. Established techniques such as dark-field or total internal reflection microscopy can reduce the background, but they still illuminate a large area, making it difficult to eliminate unwanted scattering from dust or surface defects. Furthermore, the observation time is again limited by how quickly the nanoparticles diffuse out of the illuminated volume. Alternative methods such as interferometric detection [7, 13, 14, 15, 16, 17] or measuring optical absorption with photothermal microscopy [13] can detect the elastic scattering from nanoscale objects such as single viruses by measuring the deviations from background scattering induced by the presence of the object. However, if the background varies over time, the fluctuations in background can overwhelm the signal from the detected objects. Hence in these differential schemes the objects are typically immobilized, confined to a surface, or labeled with a strongly scattering metal particle [14, 15]. These constraints make these methods less suitable for studying processes such as nucleation and growth in their natural three-dimensional environment.

Here we present a technique that can track mobile, unlabeled nanoparticles and macromolecules over a wide range of time scales, from microseconds to tens of second, and in principle for much longer. The key element of our technique is a single-mode, step-index optical fiber with an open nanoscale channel inside the high-index core (Fig. 1a,b). This channel is filled with a liquid containing the nanoparticles. Light is then guided through the fiber's core. Because the channel diameter is smaller than the wavelength of light, a portion of the optical mode in the fiber overlaps with the nanochannel cross section as depicted in Fig. 1c,d. The guided light scatters off the nanoparticles, travels through the transparent fiber cladding, and is collected by a microscope objective mounted perpendicular to the fiber axis.

This arrangement has three key features that enable simultaneously high-speed and long-duration measurements of nanoparticle dynamics. First, the orthogonal arrangement of illumination and detection directions efficiently separates the two and provides high signal-to-background ratio, obviating the need to immobilize the nanoparticles. Furthermore, the drawn silica fiber has an extremely low residual background scattering that is static and can easily be subtracted. Second, the fiber prevents the particles from diffusing out of the depth of focus of the objective. The investigated particles are continuously illuminated within the imaging volume, while their Brownian motion remains, for all practical purposes, unconstrained – albeit with hydrodynamic corrections due to the walls of the channel. Third, the coherence of the elastically scattered light enables interferometric measurements of distances between particles.

We show that the technique can track multiple dielectric particles as small as 19 nm in diameter at frame rates up to 3.5 kHz, the maximum allowed by our detector. Since we use elastic scattering, for any value of the exposure time the in-coupled light intensity can be adjusted to fill the full electronic dynamic range of the sensor, so much higher speeds are possible. To demonstrate the usefulness of the technique for biophysics experiments and diagnostic applications, we track unlabeled, freely diffusing, single cowpea chlorotic mottle virus (CCMV) virions with a molecular weight of only 4.6 megadaltons. To our knowledge, this is the smallest virus to be tracked at the single virion level with optical scattering. While CCMV is pathogenic only in plants, it is representative of a large number of positive-sense single-stranded RNA viruses that infect animals and humans. Additionally, outside of its natural host [16], CCMV and its close relative brome mosaic virus have been exploited in the fabrication of novel nanomaterials [17], nanoreactors [18], and delivery vectors for small molecules [19] and genes [20]. Our method has the potential to probe the various dynamical processes both within the viral life-cycle and within the broader field of viral nanoscience.

**Results**

*Background scattering*

Our nano-fluidic step index fiber consists of a hollow cylinder (inner channel diameter 250 nm, outer dielectric core diameter 3 µm; see Fig. 1b) made of high-index doped silica, surrounded by a concentric fused-silica fiber with a slightly lower refractive index (for details of the fabrication procedure see online methods). We design the fiber with a sub-wavelength channel so that the guided optical mode has a large fraction of its field within the channel, even though it has a lower refractive index than silica. The structure therefore allows strong guided illumination of an aqueous sample. In the experiments detailed below, laser light is coupled into one end of the fiber, and the nanoparticle suspension is pulled into the opposite end by capillary forces. We collect the scattered light and image the channel with a 60x 0.95-NA oil-immersion objective placed perpendicular to the fiber (Fig. 1a). To minimize cylindrical lensing by the outer interface of the fiber, we immerse the fiber cladding in index-matching oil on top of a flat glass slide. Images of the nanoparticles are captured by a high-speed sCMOS camera.

The background scattering in our nano-fluidic fiber is several orders of magnitude lower than in other opto-fluidic devices used for particle detection, such as lithographically fabricated waveguides [21]. The propagation loss of the fundamental core mode is smaller than 50 dB/km in the visible range; equivalently, only one part in $10^8$ of the in-coupled power is lost per micrometer of propagation length along the fiber (see online methods). The loss, which is primarily due to scattering from different material interfaces, generates a stationary background speckle in our images. Though the loss is higher than in commercially available optical fibers, the resulting background speckle is weak – comparable to scattering from a 15-nm dielectric nanoparticle. Because this background is stationary, it can be removed through image processing. Furthermore, we anticipate that the fiber fabrication process could be optimized to achieve even lower losses. Other silica-based fibers such as photonic-crystal fibers can in principle guide both light and fluids in the core [22], but the array of holes surrounding the core obstructs the image, preventing accurate particle localization and tracking.

*Tracking weakly scattering nanoparticles*

Using the nano-fluidic optical fiber, we are able to image and track freely diffusing polystyrene nanoparticles as small as 19 nm. Although the size of these dielectric particles is similar to that of gold nanoparticles used in other tracking experiments [23, 24], their scattering cross-section is about 1000 times smaller, owing to much smaller refractive index contrast and absence of plasmonic enhancement. Nonetheless we are able to image them at frame rates of 1 kHz under a guided optical power of less than 10 mW. A single frame of raw data showing several particles of different sizes in a mixture of particles of diameters 19, 35, and 51 nm (sizes reported by the manufacturers) is shown in Fig. 1e-g. Within each such frame we are able to capture multiple particles with scattering intensities spanning two orders of magnitude.

To identify particles within the mixture, we localize their positions from the images and construct tracks using custom-developed software (see online methods). An example time-trace obtained by this procedure is shown in Fig. 2a. We can distinguish different particles in the mixture by their intensities and – after analyzing their tracks (Figure S3 in online methods) – by their diffusion coefficients. To obtain more accurate tracks for the smallest particles, we use a suspension containing only 19-nm particles and triple the illumination power. For each track we calculate the diffusion coefficient and the average scattered intensity, normalized to the average background scattering. The diffusion coefficient and scattered intensity distributions are plotted as a function of the specified diameters of the particles in Figure 2b-c.

The data shown in these plots allow us to quantify the precision and accuracy of our method. We first examine the precision. The variations in particle size as extracted from the diffusion coefficients seen in Fig. 2b are larger than those extracted from the measured intensities. This difference can be traced to the uncertainty in localizing the position of the particles. We quantify the localization error by examining the measured displacements. For the 35- and 51-nanometer-size particles, the displacement distributions are Gaussian at all lag times, and the mean-square displacements (MSD) for about 95 percent of our trajectories scale linearly with lag time down to the smallest measured lag time (1 ms). For the 19-nm particles, the MSDs are linear down to about 5 ms. By considering the largest positive intercepts in the plot of MSD as a function of lag time, we estimate a localization error of less than 50 nm for the larger particles and 300 nm for the 19-nm particles. These errors are dominated by dynamical tracking errors stemming from the blurring of the point spread function as the particles diffuse during the 1 ms exposure. The dynamic errors could be further improved by using a higher-speed camera or pulsed illumination and a collection objective with a larger numerical aperture. Although our setup was not optimized to reduce the dynamic error, in all cases the localization precision is sub-wavelength.

To characterize the accuracy, we examine the diffusion coefficients obtained from the displacements. Owing to the hydrodynamic resistance of the walls of the capillary [25, 26], the measured diffusion coefficients are smaller than the ones measured by dynamic light scattering in the bulk. However, the measured values are in good agreement with those predicted by a model (yellow dashed line in Fig. 2b) of hindered diffusion in a tight capillary [25]. The model has no fit parameters. The agreement with expectations from hydrodynamics attests to the accuracy of the technique.

We foresee several possible analytical and biophysical applications for our method: characterizing the concentration and size distribution of nanoparticles in suspension, measuring the dynamics of small protein aggregates such as viruses, and measuring the interactions between nanoparticles or biomolecules. In what follows, we describe the results of three sets of experiments showing that the technique can acquire the key information required for each of these applications.

*Characterizing nanoparticles*

First, we show that our method can be used to characterize the size distribution of nanoparticles in small fluid volumes. This is a common challenge in nanotechnology, nanomedicine [27] and clinical diagnostics [28]. Our method yields two independent measures of the particle size: the diffusivity, which can be used to determine the hydrodynamic radius, and the scattering intensity, which can be used to find the optical diameter, if the particle shape and refractive index are known. The error in the particle size obtained from the diffusivity depends on the length of available tracks – the longer the duration of the track, the more accurate is the estimate of the diffusivity. As in dynamic light scattering, the principal advantage of measuring the size through the diffusivity is that the refractive index need not be known.

If, however, the index is known, then the particle-size distribution can be more accurately evaluated from the scattered intensity, which scales as the sixth power of the particle size (Fig. 2c). This measurement is akin to static light scattering, except that it can be done on individual particles. The measured intensity for each particle fluctuates in time, owing to the non-uniform intensity distribution of the mode in the fiber (see supplementary materials and Fig. S5). These fluctuations are however averaged out for exposure times longer than 10 ms. We find that the width of the intensity histogram for a single particle, measured over an entire trajectory, is smaller than the width of the histogram of an ensemble of particles from the same batch (Fig. 2d). Hence even narrow size distributions can be characterized. From the intensity histograms shown in Fig. 2d, we calculate the mean diameter and Coefficient of Variation (CV), the ratio of the standard deviation to the mean, in three different batches of polystyrene nanoparticles. From our tracking measurements we estimated the mean sizes and polydispersity of the three particle batches to be 52.2 nm with CV of 6%, 39.6 nm with CV of 11.5% and 23.5 nm with CV of 31%. The values for the CV are reasonable, lying between those reported by the manufacturer and those measured by dynamic light scattering in bulk, and the diameters are in good agreement with both data sets.

*Tracking small RNA-virus particles*

Second, we show that the method has sufficient sensitivity to follow the dynamics of small protein aggregates. We demonstrate this by measuring the motion of individual wild-type CCMV virions (Fig. 3) that are freely diffusing in water. The hydrodynamic diameter of these particles is $26\pm5$ nm, as measured by dynamic light scattering, and their scattering cross-section is $2.3\times10^{-3}$ nm$^2$, as estimated from the mass of each virion, assuming an average refractive index of 1.45 for protein. Because this cross-section is smaller than that of the 19-nm polystyrene nanospheres, detecting and tracking the virions is a challenge: we find that the scattering signal from a single virion in a channel with diameter of 400 nm is about 0.3 times the average background scattering in the fiber. Note that the channel diameter for this experiment is larger than that used for the polystyrene nanospheres, and hence the average mode intensity inside the channel is smaller. However, after background subtraction, we obtain a signal to background ratio (SBR) of 10, which is sufficient to track individual virions for tens of seconds (Fig 3b).

We measure an average diffusion coefficient of $15.0 \pm 3.6$ μm$^2$/s, which agrees with the theoretical value of 13.5 μm$^2$/s calculated using the hindered diffusion model [25] for 26-nm spheres in a 400-nm cylindrical channel. To further verify that we are indeed tracking the virus, we examine the polydispersity, as determined by the distribution of the detected scattered intensity. We find an upper bound on the mass polydispersity of 7.4% (see supplementary materials), consistent with the narrow size distribution expected of the virions. To our knowledge, these are the smallest viruses that have been tracked in real space using elastic light scattering.

*Measuring particle-particle interactions from interference of scattered light*

Third and finally, we show that the method can be used to resolve the distance between two diffusing nanoparticles that are within a wavelength of one another and measure their interactions. To make this measurement we take advantage of the coherence of elastic light scattering. In coherent scattering, the total intensity of two particles that are closer than roughly half the wavelength to one another is greater than the sum of the intensities of two isolated particles, owing to constructive interference. The nonlinear superposition of intensities in elastic scattering enables instantaneous distance measurements that are not possible in fluorescence or incoherent scattering techniques, which rely on time-averaged emission rates.

We measure the interaction between two particles whose tracks cross (Fig. 4a,b). For ease of analysis, we choose a case where one of the particles is temporarily stuck to the wall of the core (Fig. 4a). We define the enhancement of the intensity at the location of the immobile particle as $F = (I_{1+2} - I_1)/I_2$, where $I_{1+2}$ is the measured intensity, $I_1$ the scattered intensities from the immobile particle and $I_2$ that of the mobile particle. As can be seen in Fig. 4b, the measured intensity of the moving nanoparticle is enhanced by factors between 1.6 and 2.7 when it passes the fixed particle (note that the enhancement factor, which is proportional to $\sqrt{I_1/I_2}$, can be even larger if the enhancing particle is brighter than the probed one). By considering the measurement geometry and the actual mode profile, we can map the enhancement factor onto the separation between the two particles. This mapping works within a limited but well-defined uncertainty interval for distances smaller than 300 nm. The mapping and the attributed statistical uncertainty are shown in Fig. 4c (for details of the analysis, see the supplementary materials).

The relation between enhancement and separation allows us to measure the distribution of interparticle distances (Fig. 4d) by looking only at the intensity fluctuations. It is possible to make this measurement even when it is not possible to localize the individual particles. The distances are calculated by collecting all the data points in which the enhancement is larger than 1.5 – for which the value of the enhancement maps onto a narrow, monomodal distribution of distances.

Assuming Boltzmann statistics, we then calculate the interaction potential as a function of distance between the two particles (Fig. 4e). The measured interaction potential is consistent with expectations for two particles interacting through a short-range electrostatic repulsion: it is flat at distances larger than the Debye length, which we estimate to be 3 nm, and the variations are small compared to the thermal energy $k_BT$.

**Discussion**

We have shown that the opto-fluidic fiber platform enables the simultaneous measurement of the hydrodynamics and optical properties of unlabeled, freely diffusing nanoparticles with subwavelength precision and microsecond time resolution. Here we consider the applications for this technique, and we discuss the limits to sensitivity and temporal resolution. First, we note that there are some applications that are not well-suited for the method; in particular, it would be difficult to measure dynamics in living cells or cell extracts, where the large density of scatterers would likely overwhelm the signal from the particles of interest. Fluorescent techniques have an advantage for such applications: they can track labeled particles on a dark background even in dense media, albeit with low temporal resolution.

A more natural application for our method is the study of biochemical processes that occur *in vitro*. The fiber may be ideal for probing the nucleation and growth of biomolecular aggregates such as fibrils or viruses. In such systems, nucleation may take a long time to occur but growth can proceed much more rapidly. Our method is well-suited to studying such processes because it can probe a wide range of timescales, from microseconds to many seconds (and longer). It would be particularly interesting to follow the self-assembly of individual viral capsids from their constituent proteins and nucleic acids. This key step in the viral life cycle is thought to follow a nucleation-and-growth pathway [29], but direct experimental evidence for the assembly mechanism is scarce.

There are further applications in medical diagnostics and colloidal science. Diagnostic applications could take advantage of the broadband transmission of the fiber, which enables multi-color spectroscopic measurements at high spatial and temporal resolution. Simultaneous spectroscopy and tracking could detect specific pathogens or disease markers, such as extracellular vesicles [30], in bodily fluids. Applications in colloidal science include characterizing single particles and measuring dynamics. The fiber geometry allows one to measure scattering at multiple angles and polarizations simultaneously, enabling high-precision, single-particle dynamic and static light scattering measurements. Such experiments could quantify the shape and rotational diffusion [31] of non-spherical particles or aggregates. The platform could also be a useful tool for controlled measurements of electrostatic and electrokinetic effects in transport of charged particles through nanopores [32]. By combining particle tracking with capillary electrophoresis it may even be possible to study the charging dynamic of proteins [33] at a single-particle level [34] in biologically relevant environments.

Some of these applications will require improving the sensitivity and temporal resolution of the device. The sensitivity is currently limited by background scattering, which can be reduced in two ways. First, the fiber design and fabrication can be optimized; the current fibers are prototypes, and there are several parameters that can be varied in later production runs to make them smoother and more homogeneous. Second, one can use a low-coherence light source or rapidly modulate the laser current to reduce the background speckle [35]. The temporal resolution in our experiments is limited by the detector speed; the intensities we measure for our weakest scatterers – CCMV virions – are well above photon shot noise even at kilohertz frame rates. As in other techniques, the shot noise poses the ultimate limit on temporal resolution if the camera is sufficiently fast. However, because our method is based on elastic scattering, it is possible to use high illumination intensities (and, if necessary, short pulses) without thermally damaging the sample. Thus the shot-noise limit can be reduced further than in techniques that rely on inelastic scattering or fluorescence. Furthermore, the scattering of specific particles or complexes of interest can be enhanced by conjugating them to strongly-scattering nanoparticles. With both lower background and enhanced scattering, it may even be possible to measure the dynamics of single proteins.

**Acknowledgments:**

SF acknowledges fruitful discussions with Vahid Sandoghdar during an early stage of this research. YL acknowledges the support of the Pappalardo fellowship in Physics at MIT. We acknowledge the support of the Harvard MRSEC funded by NSF grant no. DMR-1420570. This work was performed in part at the Center for Nanoscale Systems (CNS), a member of the National Nanotechnology Infrastructure Network (NNIN), which is supported by NSF grant no. ECS-0335765. CNS is part of Harvard University.

**Statement of competing financial interests:**

Stefan Weidlich works for Heraeus Quarzglas GmbH & Co. KG, which is a global supplier of silica products. Within their portfolio, Heraeus Quarzglas also offers material solutions for the fiber optical industry.

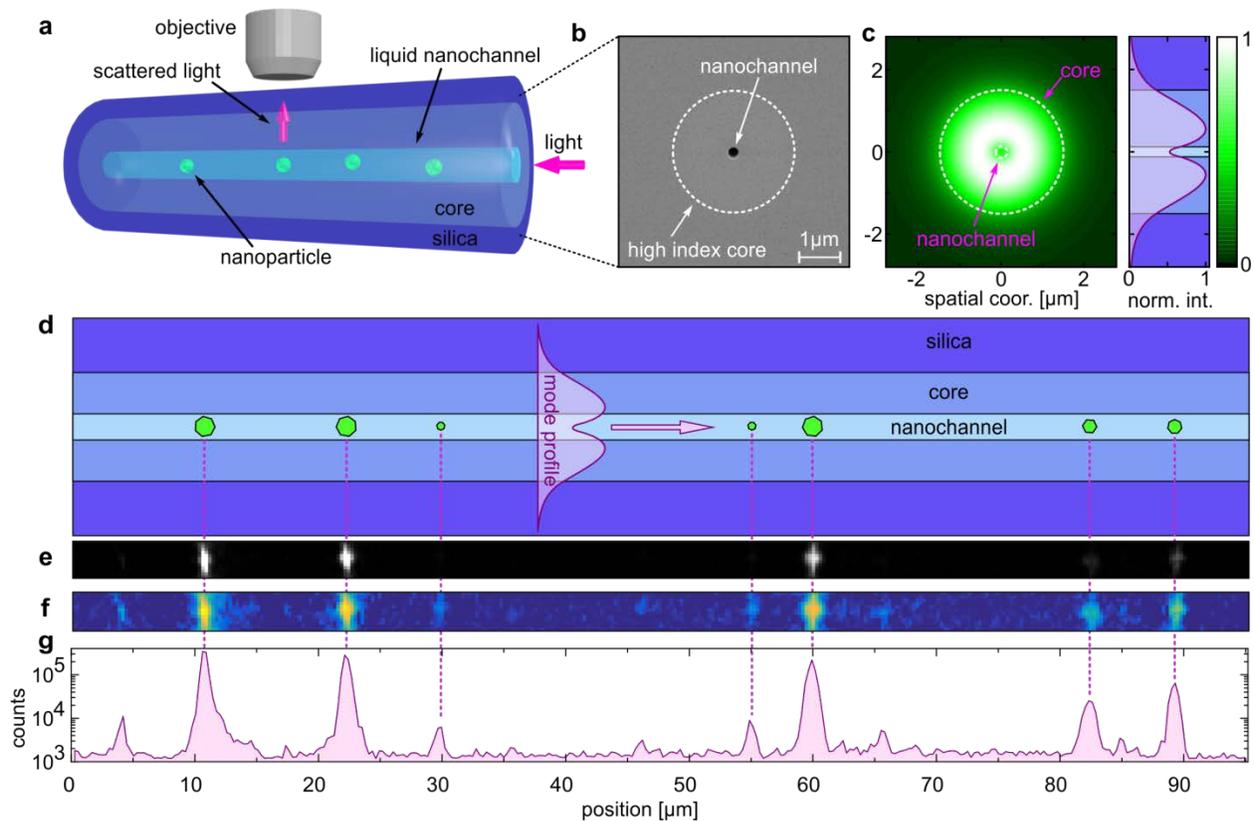

**Fig. 1**. **Detection of dielectric nanoparticles in a nano-fluidic fiber** (a) Schematic of the apparatus: light is coupled into the core of the fiber, and particles are pulled into the nanochannel by capillary forces. The scattered light is imaged through an objective oriented perpendicular to the propagation direction. (b) A scanning electron micrograph of the cross section of a fiber shows a diameter of 247±6 nm for the channel and 2.92±0.01 µm for the core. (c) The calculated intensity of the optical mode at 670 nm assuming the channel is filled with water. The plot on the right shows the profile of the mode. Blue regions correspond to those in part (d), which shows a schematic view of the fiber structure, the optical mode, and particles inside the liquid nanochannel. (e) An image of the scattered light (exposure time 1 ms) captured by the objective, with the static background speckle subtracted. The sample is a mixture of 19, 35, and 51 nm latex particles. (f) The same image shown in false-color and with a logarithmic intensity scale makes it possible to see all three sizes of particles. (g) Semi-logarithmic plot of the detected intensity as a function of position in the image.

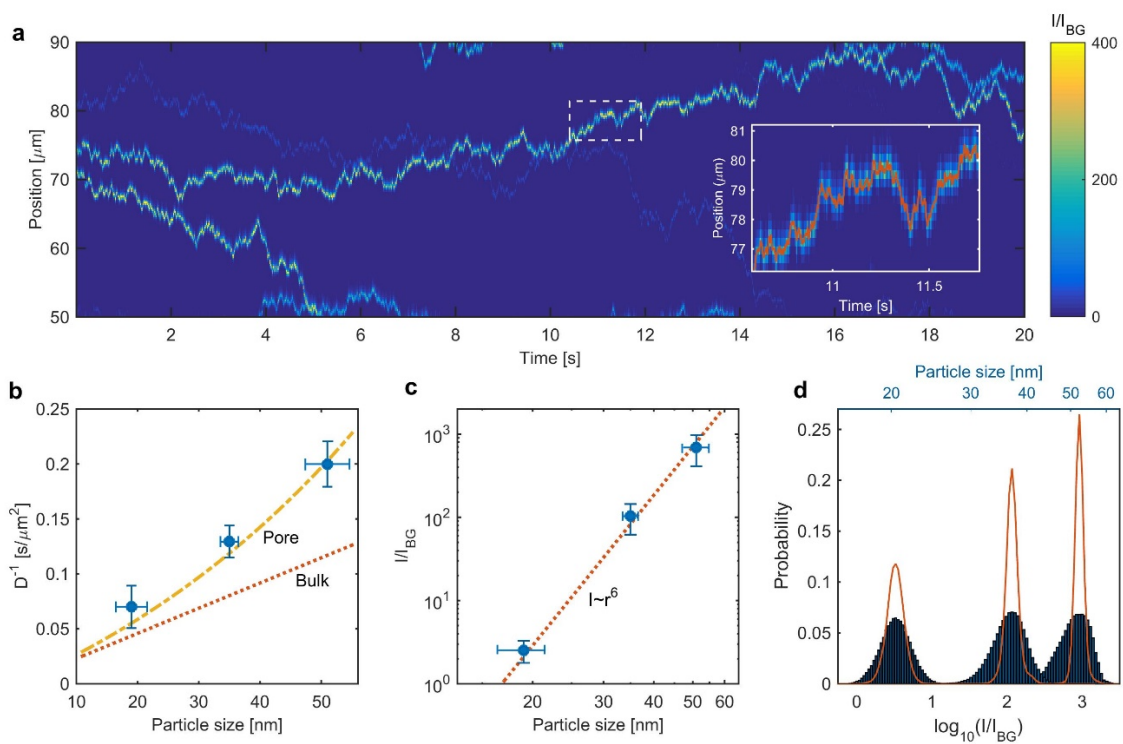

**Fig. 2**. **Tracking of unlabeled dielectric nanoparticles** (a) Axial position of polystyrene nanoparticles as a function of time, as measured by our tracking algorithm for a representative data set. Colors show the measured intensity of the particle as a function of position. The inset shows part of a track (red line) overlaid on the intensity data. (b) The inverse of the measured diffusion coefficient as a function of particle diameter (as specified by the manufacturer). Error bars represent the standard deviation of the distribution of diffusion coefficients measured for different particles of the same size. The red dashed line represents the values expected for unhindered bulk diffusion, while the yellow dashed line, calculated according to [25], takes into account the hydrodynamic drag in the capillary using no fit parameters. (c) Log-log plot of the measured normalized scattered intensity as a function of the particle diameter. The dashed line represents the Rayleigh scattering formula. (d) Histograms of the detected intensities for all particles (blue bars) and of the mean intensity of a single particle from each batch (red curves).

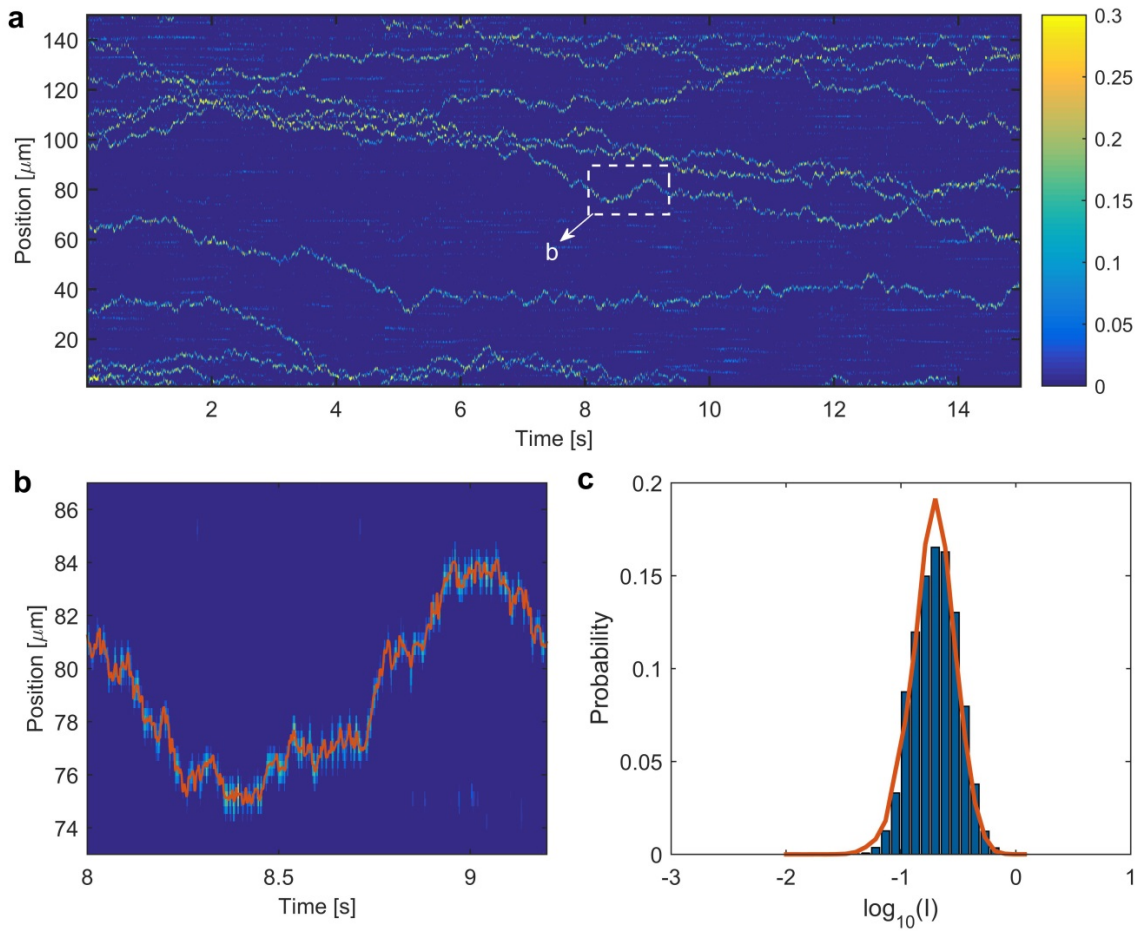

**Fig. 3**. **Tracking of single, unlabeled Cowpea Chlorotic Mottle Viruses (CCMVs).** (a) A plot of the detected tracks shown along with the measured intensities, as in Figure 2a. (b) A close-up of one of the tracks. The detected position of the particle is shown by the red line. (c) Intensity histogram for all detected CCMV virions (blue bars) and the intensity histogram of a single particle (red curve). The close fit between the two is evidence for the narrow size distribution of the virions.

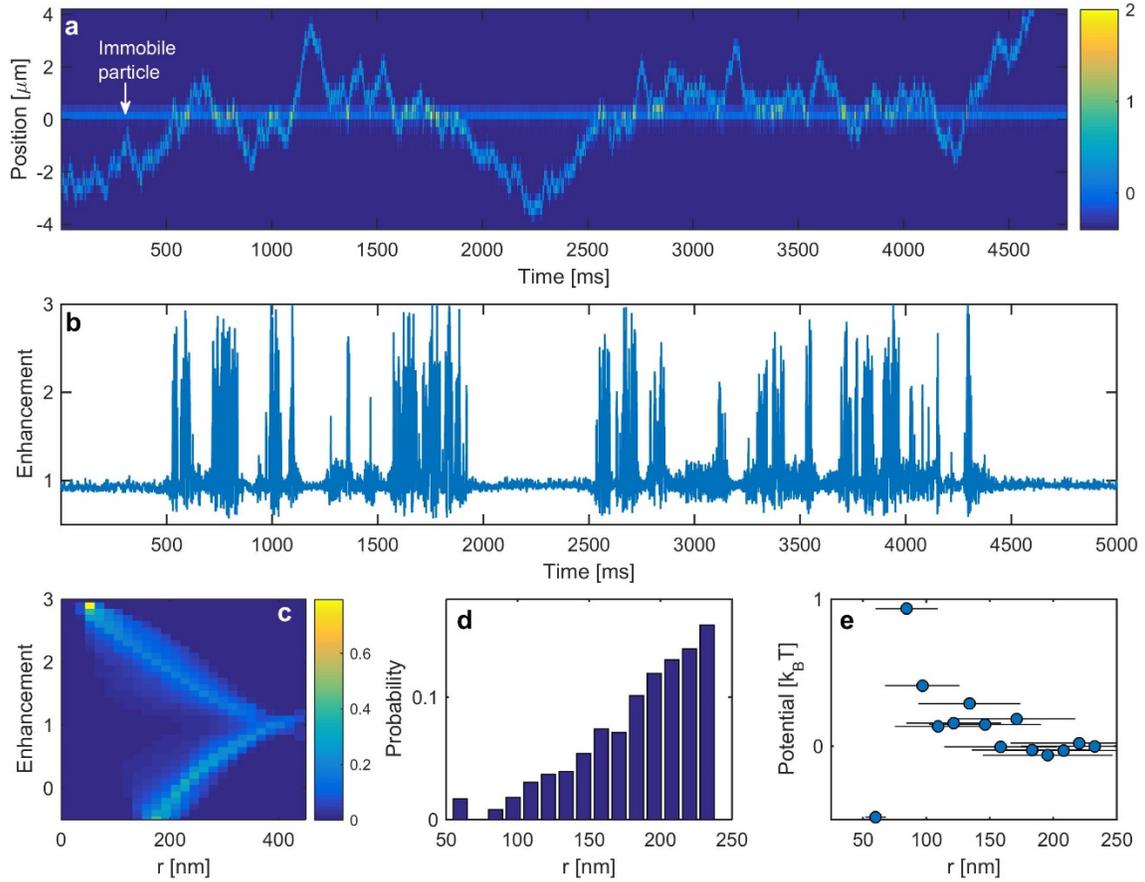

**Fig. 4**. **Interferometric enhancement.** (a) The scattering signal, measured at 3.5 kHz and shown as a function of time and position in the fiber, from a diffusing particle as it diffuses near an immobile particle. (b) The enhancement detected at the position of the immobile particle as a function of time. (c) Numerically-integrated statistical relation between enhancement and inter-particle distance $r$. The color scale depicts the calculated joint probability density distribution $P(F, r)$, assuming the position of the diffusing particle is uniformly distributed within the channel, excluding the fixed particle volume. (d) Histogram of interparticle distances, as deduced from the enhancement values using the model in (c). (e) The measured interaction potential between the particles versus the distance.

16